\documentstyle[preprint,osa]{revtex}
\begin{document}
\preprint{in press in Phys. Rev. B}
\title{
Spin fluctuations in CuGeO$_3$ probed by light scattering
}
\author{
Haruhiko Kuroe, Jun-ichi Sasaki, and Tomoyuki Sekine
}
\address{
Department of Physics, Sophia University,\\
7-1 Kioi-cho, Chiyoda-ku, Tokyo 102, Japan
}
\author{ Naoki Koide, Yoshitaka Sasago, and Kunimitsu Uchinokura
}
\address{
Department of Applied Physics, The University of Tokyo,\\
7-3-1 Hongo, Bunkyo-ku, Tokyo 113, Japan
}
\author{Masashi Hase}
\address{The Institute of Physical and Chemical Research (RIKEN),\\
 2-1 Hirosawa, Wako-shi, Saitama 351-01, Japan}
\maketitle
\begin{abstract}
	We have measured temperature dependence of low-frequency Raman spectra in
CuGeO$_3$, and have observed the quasi-elastic scattering in the $(c,c)$
polarization above the spin-Peierls transition temperature.
	We attribute it to the fluctuations of energy density in the spin system.
The magnetic specific heat and an inverse of the magnetic correlation length
can be derived from the quasi-elastic scattering.
	The inverse of the magnetic correlation length is proportional to
$(T-T_{SP})^{1/2}$ at high temperatures.
	We compare the specific heat with a competing-$J$ model.
	This model cannot explain quantitatively both the specific heat and the
magnetic susceptibility with the same parameters.
	The origin of this discrepancy is discussed.
\end{abstract}
\pacs{PACS numbers: 78.30.-j, 78.20.-e, 75.90.+w}
%78.30.-j : Infrared and Raman spectra and scattering
%78.20.-e : Optical properties of bulk materials
%75.90.+w : Other topics in magnetic properties and materials

\section{INTRODUCTION}

	Hase, Terasaki, and Uchinokura\cite{Hase} reported that the magnetic
susceptibility in all directions in CuGeO$_3$ decreases exponentially below
$T_{SP}$ = 14 K, and they concluded that it is due to a spin-Peierls (SP)
transition.
	In the SP phase, an energy gap between the singlet ground and the triplet
excited states opens and a lattice dimerization is formed.\cite{Pytte}
	After this report, many experimental studies have been performed in this
compound to reveal the nature of the SP phase.
	The magnetic excitations and the SP gap have been extensively studied by
inelastic-neutron\cite{netronbynishi} and Raman scattering.\cite{RamanKuroe}
	Moreover the lattice dimerization was observed below $T_{SP}$,
\cite{latticedimerization,Pouget,latticedimerization2} and then the SP
transition was established in CuGeO$_3$.

	On the other hand, Hase, Terasaki, and Uchinokura\cite{Hase} pointed out that
the magnetic susceptibility above $T_{SP}$ cannot be described quantitatively
by Bonner-Fisher's theory\cite{BonnerFisher} based on an $S=1/2$
one-dimensional (1D) Heisenberg antiferromagnetic (AF) Hamiltonian with a
nearest-neighbor (nn) exchange interaction.
	Recently, Riera and Dobry\cite{Riera} and Castilla, Chakravarty, and
Emery\cite{Castilla} proposed independently for CuGeO$_3$ an $S=1/2$ 1D
Heisenberg model with the nearest- and next-nearest-neighbor (nnn) AF exchange
interactions in the chain (a competing-$J$ model).
	The Hamiltonian of this model is written as
\begin{equation}
{\cal H} = J \sum_{i=1}^{N}
           \left({\bf S}_i \cdot {\bf S}_{i+1}
           + \alpha {\bf S}_i \cdot {\bf S}_{i+2} \right)\ ,
\label{Hamiltonian}
\end{equation}
	where $J$ $(>0)$ and $\alpha$ $(>0)$ are the nn AF exchange interaction and
the ratio of the nnn interaction to the nn one, respectively.
	In this model, if $\alpha$ is larger than a critical value $\alpha_c$, an
energy gap between the singlet ground and the triplet excited states appears
without any lattice dimerization.
	$\alpha_c$ is estimated to be 0.2411.\cite{Tonegawa,nnngap}
	Riera and Dobry suggested that the model with $J=160$ K and $\alpha=0.36$ can
explain the magnetic susceptibility above $T_{SP}$.\cite{Riera}
	On the other hand, Castilla, Chakrabarty, and Emery reported that the model
with $J=150$ K and $\alpha=0.24$ can explain both the magnetic susceptibility
above $T_{SP}$ and a dispersion of magnetic excitations at low
temperatures.\cite{Castilla}

	In order to check the validity of this model, other experimental results will
be compared with this model. As is well known, the magnetic specific heat
($C_m$), as well as the magnetic susceptibility, is an important physical
quantity.
	Therefore, it is necessary to measure accurately $C_m$ at high temperatures.
	However, $C_m$ at high temperatures has been poorly studied, although the
anomalous specific heat around $T_{SP}$ in CuGeO$_3$ was investigated in many
experiments.\cite{specificheat,Kobayashi,Sahling,Kuo,Liu,Oseroff,Weiden}
	Weiden {\it et al.}\cite{Weiden} observed the specific heat below 300 K in
polycrystalline CuGeO$_3$ using a continuous-heating adiabatic method.
	They obtained $C_m$ after subtracting a calculated lattice specific heat from
the observed one.
	The lattice specific heat was assumed to be written as a sum of Debye
functions.
	They reported that $C_m$ shows a broad maximum around 40 K.
	As we will describe below, our result contradicts theirs.
	As is well known, $C_m$ at high temperatures cannot be obtained accurately by
this method because effects of optical phonons, which play an important role in
the specific heat at high temperatures, were not taken into account precisely.

	It is noted that the quasi-elastic scattering is a powerful tool to derive
precisely $C_m$ especially at high temperatures.
	In our brief report,\cite{KuroeRaman2} a large tail near an incident laser
line with a strong temperature dependence was observed at high temperatures in
single-crystal CuGeO$_3$, the profile in the Stokes and anti-Stokes scattering
was fit to a Lorentzian curve.
	We assigned it to the quasi-elastic scattering due to spin-energy
fluctuations.
	We can obtain roughly the magnetic specific heat from the quasi-elastic Raman
scattering in CuGeO$_3$, because the integrated intensity of the quasi-elastic
scattering is proportional to $C_mT^2$ where $T$ is the
temperature.\cite{Halley}
	The quasi-elastic scattering was applied to some magnets such as a
two-dimensional antiferromagnet FePS$_3$,\cite{FePS3} and later a 1D
antiferromagnet KCuF$_3.$\cite{KCuF3}

	In our brief report\cite{KuroeRaman2} we observed Raman spectra in both the
Stokes and anti-Stokes regions at seven different temperatures.
	The Raman spectra, however, contained a weak component of the direct
scattering, judging from the fact that a plasma line of the incident laser line
appeared at -13.4 cm$^{-1}$.
	The quality of the sample was probably worse than that used in the present
work.
	It is, of course, necessary to use high-quality samples in order to the
effects of the direct scattering when we measure the quasi-elastic scattering.
	Then we did not obtain accurately the integrated intensity of the
quasi-elastic scattering, in particular at low temperatures.
	At low temperatures the intensity of the quasi-elastic scattering decreased in
CuGeO$_3$ and it was considerably affected by the direct scattering, when we
estimated the integrated intensity in our brief report.

	Very recently, on the other hand, van Loosdreht {\it et al.}\cite{Loosdreht}
measured Stokes Raman spectra with a backscattering geometry and proposed that
the low-frequency Raman spectrum at high temperatures originated from the
diffusive-type fluctuations in the four-spin time correlation functions.
	Richards and Brya\cite{Richards} theoretically studied Raman scattering from
the diffusive-type fluctuations in the spin system by using a four-spin time
correlation function and obtained a result that the profile was described as a
Gaussian-like line at $\omega$=0 at high temperatures.
	Therefore the model proposed by van Loosdreht {\it et al}. seems to contradict
the interpretation for the quasi-elastic scattering in our brief
report.\cite{KuroeRaman2}

	Therefore the careful and precise measurement of the low-frequency Raman
spectrum in both the Stokes and anti-Stokes regions is needed in a high-quality
sample in order to reveal clearly the origin of the quasi-elastic Raman
scattering.
	In the present paper we shall obtain the magnetic specific heat precisely
above $T_{SP}$ and study effects of the nnn AF exchange interaction along the
magnetic chain on the magnetic properties of CuGeO$_3$.

\section{EXPERIMENTAL DETAILS}
	A single crystal used in this study was made by a floating-zone method.
	The dimension of the crystal is $1.2 \times 1.8 \times 10.7$\,mm$^3$ parallel
to the $a$, $b$, and $c$ axes, respectively.\cite{notation}
	The single crystal has flat $ac$ and $bc$ faces.
	On the other hand, it is difficult to polish the $ab$ face because of cleavage
on the $bc$ face.
	However, effects of local strain on the $ab$ face were negligible when the
light was incident and scattered on the $ac$ and $bc$ faces.
	The $bc$ face was attached to the sample holder by dilute varnish.

	Raman-scattering experiments were carried out using the 5145-\AA \ line of an
Ar$^+$ ion laser.
	The polarized quasi-elastic Raman scattering was measured in either a
right-angle or a quasi-back scattering geometry.
	The detailed temperature dependence of the quasi-elastic Raman scattering was
measured in the right-angle scattering geometry.
	To eliminate plasma lines and to get a complete linear polarization, we made
the laser light pass through a filter and a polarizer and then the laser light
was incident on the sample in a cryostat.
	The light was incident on the $ac$ face for the right-angle scattering and on
the $bc$ face for the quasi-back scattering.
	The scattered light was dispersed by a Jobin-Yvon U1000 double-grating
monochromator and was detected by a photon counting system.
	The whole system is controlled by a microcomputer.
	To measure Raman spectra below room temperature, we used a closed-cycle
cryostat.
	The temperature of the sample is controlled by a PID temperature controller.
	The temperature of the sample is equal to that of the sample holder within
$\pm$0.1 K.
	Since the sample is transparent, the local temperature rise due to the
incident light is negligible.

\section{THEORIES}
\subsection{QUASI-ELASTIC SCATTERING}
	We summarize briefly the theory of the quasi-elastic light scattering in
antiferromagnets.
	According to Halley's theory,\cite{Halley} the intensity of the quasi-elastic
light scattering is given by the following Fourier component of the correlation
function of the magnetic energy density:
\begin{equation}
I(\omega) =
\gamma
\int_{-\infty}^{\infty}
dt
e^{-i\omega t}
\langle
E({\bf q},t)E^{*}(-{\bf q},0)
\rangle
\ ,
\label{thermalaverage}
\end{equation}
	where $E({\bf q}, t)$ and $\hbar \omega$ denote the magnetic energy density
with the scattering vector ${\bf q}$ and the energy transfer, respectively.
$\gamma$ is the ${\bf q}$- and $\omega$-independent but polarization-dependent
coefficient. We assume that $\gamma$ is temperature-independent.
	$\langle \cdots \rangle$ represents a thermal average.
	In the hydrodynamic condition,\cite{Halperin} Eq. (\ref{thermalaverage}) is
transformed to
\begin{equation}
I(\omega ) =
\frac{\gamma \omega}{1-e^{-\hbar \omega/k_B T}}
\frac{C_m T D_T q^2}{\omega^2 + (D_T q^2)^2}\ .
\label{lowtemperature}
\end{equation}
	Here $D_T$ is a thermal diffusion constant which is given by
\begin{equation}
	D_T=K/C_m \ ,
\label{thermaldiffusion}
\end{equation}
where $K$ is the magnetic contribution to the thermal conductivity.
	$q$ is approximately equal to $|{\bf q_0}|\sin{\theta/2}$, where ${\bf q_0}$
and $\theta$ are the wave vector of the incident light and the scattering
angle, respectively.
	In a high-temperature approximation, Eq. (\ref{lowtemperature}) can be written
as the following Lorentzian curve
\begin{equation}
I(\omega) =
\gamma'
\frac{C_m T^2 D_T q^2}{\omega^2 + (D_T q^2)^2}\ ,
\label{hightemperature}
\end{equation}
	where $\gamma'  = k_B \gamma / \hbar$.
	Using Eq. (\ref{hightemperature}), we can estimate the magnetic specific heat
and the thermal diffusion constant from the integrated intensity and the half
width at half maxima of the quasi-elastic scattering, respectively.

\subsection{MAGNETIC SPECIFIC HEAT OF SPIN CHAIN WITH COMPETING EXCHANGE
INTERACTIONS}

	We can calculate the magnetic specific heat using all energy levels in the
spin system.
	Bonner and Fisher calculated magnetic susceptibility and magnetic specific
heat in the spin system whose Hamiltonian is expressed by Eq.
(\ref{Hamiltonian}) with $\alpha$ = 0 and 3 $\leq$ $N$ $\leq$
11.\cite{BonnerFisher}
	They calculated all energy levels under the periodic condition by means of an
exact diagonalization.
	Because a finite-size effect appears at low temperatures, physical quantities
at $N \rightarrow \infty$ were estimated from an extrapolation.
	The magnetic specific heat of an infinite 1D AF chain with the nn AF exchange
interaction has a broad maximum at $k_BT \approx 0.481 J$.

	In the case of $\alpha = 0.5$ (Majumder-Ghosh model), the ground state of the
system with $4 \leq N \leq 8$ was studied analytically, and was revealed to be
products of dimer states with double degeneracy.\cite{Majumder}
	This model was developed to the case of $N \rightarrow \infty$ and higher
$S$.\cite{Affleck}
	An energy gap between the singlet ground and the triplet excited states opens
in this case, while the magnetic excitation is gapless for $\alpha = 0$.
	Therefore, a phase transition takes place at $0 < \alpha < 1/2$.
	Tonegawa and Harada calculated the energy gap between the singlet ground and
the triplet excited states numerically when $N \leq 20$,\cite{Tonegawa} and
after this work, Okamoto and Nomura found that $\alpha_c$ at $N \rightarrow
\infty$ is 0.2411 from their calculation when $N \leq 20$.\cite{nnngap}

	We calculated all energy levels by means of the exact diagonalization for
various $\alpha$ when $4 \leq N \leq 14$.
	We used a standard subroutine of the Householder method for a numerical
diagonalization.
	We calculated the magnetic specific heat ($C_{m,N}$) for $N$ spins and
investigated $D_N=[NC_{m,N}+(N-1)C_{m,N-1}]/(2N-1)$ vs. $N$.
	$D_N$ almost converges a value ({\it i.e.}, $C_m$) up to $N=14$ at least when
$T>0.5 T_{max}$.
	Here $T_{max}$ is the temperature at which $C_m$ has a maximum value
$C_{m,max}$.
	Therefore we show $D_N$ for $N=14$ as $C_m$ in Fig. 1a.
	$T_{max}$ and $C_{m,max}$ decrease and the peak becomes broader with
increasing $\alpha$.
	We also calculate $C_m/C_{m.max}$ vs. $T/T_{max}$ to obtain $J$ and $\alpha$
from our observed data, and it is shown in Fig. 1b.

\section{RESULTS AND DISCUSSION}

	Figure 2a shows polarized Raman spectra of CuGeO$_3$ at room temperature
between -100 and 100 cm$^{-1}$ strongly in the right-angle scattering
geometries.
	The Raman spectra in the regions of $\omega > 0$ and $\omega < 0$ correspond
to the Stokes and the anti-Stokes components, respectively.
	Since the intensity of the direct scattering is very strong, we cannot obtain
Raman spectra in the region of $|\omega| < 7$ cm$^{-1}$.
	The quasi-elastic scattering is strongly observed in the $(c,c)$ polarization,
while it is not observed in the offdiagonal geometries, i.e., the $(a,b)$,
$(a,c)$, and $(c,b)$ ones.
	In the SP phase, we observed two-magnetic-exciton Raman spectrum below 250
cm$^{-1}$ strongly in the $(c,c)$ polarization.\cite{RamanKuroe}
	Thus, the magnetic fluctuations should appear strongly in the $(c,c)$
polarization above $T_{SP}$, and we conclude that the quasi-elastic scattering
of CuGeO$_3$ originates from the spin fluctuations in the spin system.

	Figure 2b shows polarized Raman spectra between -100 and 500 cm$^{-1}$ at room
temperature in the right-angle ($b(c,c)a$) and the quasi-back
($a(c,c)\overline{a}$ and $a(c,c)\overline{a}$) scattering geometries.
	Because the Raman spectra in the $b(c,c)a$ and the $a(c,c)\overline{a}$
scattering geometries were obtained with different scattering volumes, we
cannot compare these scattering efficiencies directly.
	Then we normalized these Raman intensities by the integrated intensity of the
181-cm$^{-1}$ $A_g$ phonon mode.\cite{RamanKuroe}
	On the other hand, since the $a(b,b)\overline{a}$ Raman spectrum was measured
in the same experimental condition as the $a(c,c)\overline{a}$ spectrum, we can
compare these intensities directly.
	The 181-cm$^{-1}$ $A_g$ phonon mode may have different intensities in the
$a(b,b)\overline{a}$ and the $a(c,c)\overline{a}$ geometries.
	Two peaks at 114 and 221 cm$^{-1}$ are $B_{1g}$ modes, which are basically
forbidden in this geometry.
	One can see a larger tail near the laser line and a stronger background at
high-frequency region in the $a(c,c)\overline{a}$ geometry when compared with
those in the $b(c,c)a$ geometry.
	We also observed a plasma line of the incident laser at -13.4 cm$^{-1}$ in the
quasi-back scattering geometry, while we did not observe it in the right-angle
scattering geometry.
	These facts obviously indicate that the quasi-elastic scattering was strongly
influenced by the direct scattering in the quasi-back scattering geometry.
	In the $b(c,c)a$ geometry, one can notice a crossover between the
quasi-elastic scattering spectra and the incident laser line at $\pm$ 7
cm$^{-1}$.

	In our previous report,\cite{KuroeRaman2} we observed the plasma line at -13.4
cm$^{-1}$ even in the right-angle scattering geometry, because the sample
probably contains local strains and defects, and its quality was worse than
that used in the present work.
	Moreover, since the $ab$ face of the sample used in the previous
work\cite{KuroeRaman2} was not as-grown but cut and polished, the effects of
local strain on this surface and  of strong diffused reflections of the laser
light from this surface could not be avoided completely.
	Therefore, the Raman spectra in the previous report probably contained the
weak component of the direct scattering, and we could not obtain the accurate
integrated intensity of the quasi-elastic scattering especially at low
temperatures.

	We measured the temperature dependence of the low-frequency Raman spectra in
the right-angle scattering geometry using the high-quality sample with as-grown
$ab$ surfaces.
	We think that we have done all we could do to avoid the effects of the direct
scattering in the present measurements.

	Figure 3 shows typical spectra of the quasi-elastic scattering at 250, 120,
50, and 20 K.
	The intensity of the quasi-elastic scattering drastically decreases with
decreasing temperature, and we did not observe the quasi-elastic scattering
below $T_{SP}$.
	Since we did not observe any asymmetric spectra above 20 K, the observed
spectra are fitted to the following Lorentzian-type spectral function, instead
of Eq. (\ref{lowtemperature}), using the method of least squares:
\begin{equation}
I(\omega) = \frac{k^2\Gamma}{\omega^2+\Gamma^2}
             + {\rm background},
\label{spectralfucntion}
\end{equation}
	where $\Gamma$ and $k$ in Eq. (\ref{spectralfucntion}) are the damping
constant and the coupling coefficient, respectively.
	Comparing Eqs. (\ref{hightemperature}) and (\ref{spectralfucntion}), $C_m$ and
$D_T$ can be evaluated from $k^2$ and $\Gamma$.
	The background is assumed to be expressed as $A \omega + B$.
	In the fitting, we use the Raman spectra of $|\omega | > 9$ cm$^{-1}$, and
then we can avoid the effects of the incident laser line near $\omega \sim 0$.
	Below 40 K, we could not fit the observed data because their intensities were
very weak.

	Van Loosdreht {\it et al.} proposed that the origin of the low-frequency Raman
spectra is the diffusive-type fluctuations in the four-spin time correlation
functions.\cite{Loosdreht}
	According to Richards and Brya\cite{Richards}, this leads to a Gaussian-like
low-frequency Raman spectrum.
	Then, we tried to fit the observed spectra to the following Gaussian-type
spectral function:
\begin{equation}
I(\omega) = \frac{k_{G}^2}{\sqrt{2\pi}\Gamma_G}
		\exp{ \left( - \frac{\omega^2}{2\Gamma_G^2} \right) }
             + {\rm background},
\label{gaussian}
\end{equation}
where $k_G$ and $\Gamma_G$ corresponds to $k$ and $\Gamma$ in Eq.
(\ref{spectralfucntion}).
	As seen clearly in Fig. 3, the Lorentzian-type spectral functions could
reproduce the observed data much better than the Gaussian-type ones.
	Especially, the observed spectra near the incident laser line in the frequency
region of $|\omega| \leq 15$ cm$^{-1}$ did not fit to the Gaussian-type
spectral function.
	This result indicates that the observed quasi-elastic Raman scattering does
not originate from the diffusive-type four-spin fluctuations but from the
spin-energy fluctuations.

	We show the temperature dependence of $\Gamma$ and $k^2$ above 50 K in Figs.
4a and 4b.
	As is drawn by a solid curve in Fig. 4a, $\Gamma$ obeys $(T-T_{SP})^{1/2}$,
and the same result was obtained in our brief report.\cite{KuroeRaman2}
	It is noted that $\Gamma$ is proportional to $D_{T}$, and therefore to an
inverse of a magnetic correlation length $\xi^{-1}$ ($\Gamma \propto D_T
\propto \xi^{-1}$).\cite{correlationlength}
	Although we did not observe the quasi-elastic scattering below 40 K, $\Gamma$
seems to become zero around $T_{SP}$.
	Near the critical point, the thermal conductivity $K$ which depends on
short-range behavior of the system remains finite, while the magnetic specific
heat $C_m$ diverges.\cite{Halperin2}
	Then $D_T (= K/C_m)$ is predicted to vanish at $T_{SP}$.

	Pouget {\it et al.}\cite{Pouget} observed the structural fluctuations below
about 40 K by means of X-ray diffuse scattering and they reported that the
inverse correlation length of the structural fluctuations was proportional to
$(T-T_{SP})^{1/2}$.
	Recent neutron-scattering studies\cite{nishi,hirota} showed that the magnetic
correlation length agrees with the structural one near $T_{SP}$.
	The present result indicates that the obtained $\Gamma$ of the quasi-elastic
scattering above 50 K has the same temperature dependence.
	Then the inverse magnetic correlation length $\xi^{-1}$ is described as
$(T-T_{SP})^{1/2}$ not only just above $T_{SP}$ but also at high temperatures.

	Figure 5 shows $k^2/T^2$ above 50 K, which is proportional to $C_m$.
	As is seen in this figure, the data have some errors.
	Thus, we make smoothed data by a five point weighted mean (a solid curve), and
compare them with theories.
	These errors are mainly caused by the difficulty in the temperature-dependence
measurements.
	In the Raman-scattering measurements, it is not so easy to keep the exact
alignment, because the spot of the incident laser line on the sample is moved
by a thermal expansion of the cryostat.

	A broad maximum is seen in the smoothed curve around 90 K.
	In Fig. 5, we show two theoretical magnetic specific-heat curves with $J=237$
K and $\alpha =0.30$ and with $J=261$ K and $\alpha = 0.40$, which have maxima
at 90 K.
	These curves are normalized in order that the maxima ($C_{m,max}$) of both the
experimental and theoretical $C_m$ agree with each other.
	Since the smoothed data exist between two theoretical $C_m$, $J$ and $\alpha$
are roughly estimated at $250 \pm 15$ K and $0.35 \pm 0.05$, respectively.

	The competing-$J$ model seems to explain the $C_m$ of CuGeO$_3$.
	However, the magnetic susceptibility in Ref. (\ref{Hase}) is not in agreement
with the theoretical one with $J$ and $\alpha$ estimated in the present study.
	On the other hand, Riera and Dobry ($J = 160$ K and $\alpha =
0.36$)\cite{Riera} and Castilla, Chakravarty, and Emery ($J = 150$ K and
$\alpha=0.24$)\cite{Castilla} have reported different $J$ and $\alpha$, and the
theoretical magnetic susceptibility curves with their parameters agree with the
observed magnetic susceptibility.
	However, as is shown by the dashed and dotted curves in Fig. 5, the
corresponding theoretical magnetic specific-heat curves do not reproduce the
measured one.
	Here we set $C_{m,max}$ by the value of the observed datum at 90 K.
	Then we conclude that we cannot find out the parameters which explain the
observed magnetic specific heat and the observed magnetic susceptibility
simultaneously.
	This is a problem in the competing-$J$ model.

	Our $\alpha$ (0.35) is close to that of Riera and Dobry (0.36) and seems to be
larger than $\alpha_c$.
	This indicates that the energy gap remains finite even above $T_{SP}$.
	The energy gap above $T_{SP}$ may be as large as 0.015$J$ which was estimated
by Riera and Dobry\cite{Riera} when $\alpha = 0.36$.
	However the finite gap in the uniform state has not been observed by any
experiments.
	This is the second problem in this model.

	The competing-$J$ model is an interesting idea and is probably applicable to
CuGeO$_3$.
	However, this model cannot explain quantitatively all the experimental
results.
	The problems may be solved by considering (i) interchain exchange
interactions, (ii) temperature dependence of the exchange interactions, and
(iii) structural fluctuations of the lattice dimerization.

	These suggestions are due to the following reasons;

	(i) In CuGeO$_3$, the strength of the interchain interaction along the $b$
axis is about 10 \% of that along the $c$ axis,\cite{netronbynishi} and it
cannot be neglected.
	Riera and Koval suggested that the gap did not open in the uniform state even
in the case of $\alpha = 0.36$ when the interchain interaction exists, because
the interchain interaction changes the critical value $\alpha_c$.\cite{Riera2}

	(ii) Because the lattice constant of CuGeO$_3$ changes strongly with
decreasing
temperature,\cite{latticedimerization2,latticeconstant,latticeconstant2} $J$
might depend on the temperature.
	It leads to a substantial effect on the temperature dependence of the magnetic
susceptibility and the magnetic specific heat.
	When temperature dependence of $J$ is taken into account, both the observed
magnetic specific heat and the magnetic susceptibility might be explained
simultaneously in terms of the competing-$J$ model.

	(iii) Pouget {\it et al.}\cite{Pouget} observed the structural fluctuations of
the lattice dimerization whose inverse correlation length is proportional to
$(T-T_{SP})^{1/2}$ below about 40 K by X-ray diffuse scattering.
	Thus, the fluctuations may affect the magnetic specific heat and the
susceptibility through a strong spin-lattice coupling.

	Recently, Nakai {\it et al.}\cite{Iio} obtained the magnetic specific heat up
to 650 K by means of the birefringence.
	The differential of birefringence with respect to temperature gives a magnetic
specific heat.
	The temperature dependence of the magnetic specific heat shows a broad maximum
around 80 K.
	Their result agrees well with ours.

\section{CONCLUSION}
	We studied the quasi-elastic Raman scattering in CuGeO$_3$, which is caused by
spin-fluctuations.
	The quasi-elastic Raman scattering is observed only in the $(c,c)$
polarization.
	The line shapes of the Stokes and anti-Stokes Raman scattering were well
described by a Lorentzian curve around 0 cm$^{-1}$.
	This fact indicates that the quasi-elastic scattering originates from the
fluctuations of energy density in the spin system.
	The temperature dependence of the half width at half maxima, which is
proportional to $\xi^{-1}$, is well described as $(T-T_{SP})^{1/2}$.
	This is consistent with the results of the neutron-scattering studies.
	The temperature dependence of $k^2/T^2$, which is proportional to the magnetic
specific heat, was compared with the theoretical calculation by the
competing-$J$ model.
	The obtained $J=250 \pm 15$ K and $\alpha = 0.35 \pm 0.05$ are different from
those obtained in the magnetic susceptibility.
	We cannot find the parameters which explain the specific heat and the magnetic
susceptibility at the same time.
	We suggested the reasons of the discrepancies between the experimental results
and the competing-$J$ model.

\acknowledgments

This study is supported in part by the
Research Fellowships of Japan Society for
the Promotion of Science for Young Scientists
and by a Grant-in-Aid for Scientific Research from
the Ministry of Education, Science and Culture
of Japan.
The authors thank Mr. T. Nakai for useful discussion and sending their
unpublished data, and also thank Prof. T. Ohtsuki for helpful discussion about
theoretical calculations.

%\documentstyle[preprint,osa]{revtex}
%\documentstyle[12pt]{article}
%\begin{document}
%{\it Note added in proof}.
%\end{document}

\begin{figure}
\caption{
Temperature dependence of the calculated magnetic specific heat for
$\alpha$ = 0, 0.10, 0.20, 0.30, and 0.40 (a), and
their normalized specific heat (b).
}
\end{figure}

\begin{figure}
\caption{
Polarization characteristics of the Raman spectra between -100 and 100
cm$^{-1}$ in CuGeO$_3$ in the right-angle scattering geometry at 300 K (a).
Polarization characteristics of the Raman spectra between -100 and 500
cm$^{-1}$ (b).
The $b(c,c)a$ and $a(c,c)\overline{a}$ spectra in (b) are normalized by the
intensity of the 181-cm$^{-1}$ $A_g$ phonon modes.
Ar$^{+}$ denotes a plasma line of the Ar$^{+}$ ion laser at -13.4 cm$^{-1}$.
}
\end{figure}

\begin{figure}
\caption{
Typical spectra of the quasi-elastic scattering in CuGeO$_3$ at 250 (circles),
120 (squares), 50 (triangles), and 20 K (diamonds). The solid and dashed curves
denote Lorenzian curves and Gaussian curves, which are written in Eqs. (6) and
(7), respectively.
}
\end{figure}

\begin{figure}
\caption{Temperature dependence of $\Gamma$ (a) and $k^2$ (b) above 50 K in
CuGeO$_3$. The solid curve in (a) is proportional to $(T-T_{SP})^{1/2}$. }
\end{figure}

\begin{figure}
\caption{
Temperature dependence of $k^2/T^2$ above 50 K in CuGeO$_3$, and the smoothed
data.
This figure also includes the calculated magnetic-specific-heat curves for
$\alpha$ = 0.30 and $J$ = 237 K, for $J$ = 261 K and $\alpha$ = 0.40, for the
parameters of Riera and Dobry ($J$=160 K, $\alpha$ = 0.36) (Ref. 9), and for
those of Castilla, Chakravarty, and Emery ($J$=150 K, $\alpha$ = 0.24) (Ref.
10).
All the theoretical curves were normalized so that $C_{m,max}$ agrees with the
observed datum at 90 K (see text).}
\end{figure}


\begin{references}
\bibitem{Hase}
M. Hase, I. Terasaki, and K. Uchinokura,
{\rm Phys. Rev. Lett.} {\bf 70}, 3651 (1993).
\label{Hase}

\bibitem{Pytte}
For review, J. W. Bary, L. V. Interrante, I. S. Jacobs, and J. C. Bonner,
in {\it Extended Linear Chain Compounds}, ed. J. S. Miller
(Plenum Press, New York and London, 1983) Vol. III.

\bibitem{netronbynishi}
M. Nishi, O. Fujita, and J. Akimitsu,
{\rm Phys. Rev. B}{\bf 50}, 6508 (1994).

\bibitem{RamanKuroe}
H. Kuroe, T. Sekine, M. Hase, Y. Sasago, K. Uchinokura,
H. Kojima, I. Tanaka, and Y. Shibuya,
{\rm Phys. Rev. B}{\bf 50}, 16468 (1994).
\label{RamanKuroe}

\bibitem{latticedimerization}
O. Kamimura, M. Terauchi, M. Tanaka, O. Fujita, and J. Akimitsu,
{\rm J. Phys. Soc. Jpn.} {\bf 63}, 2467 (1994).

\bibitem{Pouget}
J. P. Pouget, L. P. Regnault, M. Ain, B. Hennion, J. P. Renard, P. Veillet,
G. Dhalenne, and A. Revcolevschi,
{\rm Phys. Rev. Lett.} {\bf 72}, 4037 (1994).

\bibitem{latticedimerization2}
K. Hirota, D. E. Cox, J. E. Lorenzo, G. Shirane,
J. M. Tranquada, M. Hase, K. Uchinokura, H. Kojima, Y. Shibuya,
and I. Tanaka,
{\rm Phys. Rev. Lett.} {\bf 73}, 736 (1994).

\bibitem{BonnerFisher} J. C. Bonner and M. E. Fisher,
{\rm Phys. Rev.} A{\bf 135}, 640 (1964).
The Hamiltonian in this paper is defined by
$2J_{BF}\sum_{i=1}^{N}({\bf S}_{i} \cdot{\bf S}_{i+1})$.
\label{BonnerFisher}

\bibitem{Riera}
J. Riera and A. Dobry,
{\rm Phys. Rev. B}{\bf 51}, 16098 (1995).

\bibitem{Castilla}
G. Castilla, S. Chakravarty, and V. J. Emery.
{\rm Phys. Rev. Lett.} {\bf 75}, 1823 (1995).

\bibitem{Tonegawa}
T. Tonegawa and I. Harada,
{\rm J. Phys. Soc. Jpn.}{\bf 56}, 2153 (1987).

\bibitem{nnngap}
K. Okamoto and K. Nomura,
{\rm Phys. Lett. A}{\bf 169}, 433 (1992).

\bibitem{specificheat}
H. Kuroe, K. Kobayashi, T. Sekine,
M. Hase, Y. Sasago, I. Terasaki, and K. Uchinokura,
{\rm J. Phys. Soc. Jpn.} {\bf 63}, 365 (1994).

\bibitem{Kobayashi}
T. C. Kobayashi, A. Koda, H. Honda, C.U. Hong,
K. Amaya, T. Asano, Y. Ajiro, M. Mekata, and T. Yoshida,
{\rm Physica B}{\bf 211}, 205 (1995).

\bibitem{Sahling}
S. Sahling, J. C. Lasjanias, P. Monceau, and A. Revcolevschi,
{\rm Solid. State. Commun.} {\bf 92}, 423 (1994).

\bibitem{Kuo}
Y. K. Kuo, E. Figueroa, and J. W. Brill,
{\rm Solid State Commun.} {\bf 94}, 385 (1995).

\bibitem{Liu}
X. Liu, J. Wosnitza, H. v. L\"{o}hneysen, and R. K. Kremer,
{\rm Z. Phys. B}{\bf 98}, 163 (1995).

\bibitem{Oseroff}
S. B. Oseroff, S-W. Cheong, B. Akatas, M. F. Hundley,
Z. Fisk, and L.W. Rupp, Jr.,
{\rm Phys. Rev. Lett.} {\bf 74}, 1450 (1995).

\bibitem{Weiden}
M. Weiden, J. K\"{o}hler, G. Sparn, M. K\"{o}ppen, M. Lang,
C. Geibel, and F. Steglich,
{\rm Z. Phys. B}{\bf 98}, 167 (1995).

\bibitem{KuroeRaman2}
H. Kuroe, J. Sasaki, T. Sekine, Y. Sasago, M. Hase,
N. Koide, K. Uchinokura, H. Kojima, I. Tanaka, and Y. Shibuya,
Physica B{\bf 219}-{\bf 220}, 104 (1996).

\bibitem{Halley}
J. W. Halley,
{\rm Phys. Rev. Lett.} {\bf 41}, 1065 (1978).

\bibitem{FePS3}
T. Sekine, M. Jouanne, C. Julien, and M. Balkanski,
{\rm Phys. Rev. B} {\bf 42}, 8382 (1990).

\bibitem{KCuF3}
I. Yamada and H. Onda,
{\rm Phys. Rev. B} {\bf 49}, 1048 (1994).

\bibitem{Loosdreht}
P. H. M. van Loosdreht, J. P. Boucher, G. Martinez, G. Dhalenne, and A.
Revcolevschi. {\rm Phys. Rev. Lett.}{\bf 76}, 311 (1996).

\bibitem{Richards}
P. M. Richards and W. J. Brya, Phys. Rev. B{\bf 9}, 3044 (1974).

\bibitem{notation}
We used the same notation ($Pbmm$) as that in
H. V\"{o}llenkle, A. Wittmann, and H. Howotny, {\rm Monatsh. Chem.} {\bf 98},
1352 (1967).
The symmetry of this compound is $Pmma$ in the standard orientation.

\bibitem{Halperin}
B. I. Halperin and P. C. Hohenberg,
{\rm Phys. Rev.} {\bf 188}, 898 (1969).

\bibitem{Majumder}
C. K. Majumder and D. K. Ghosh,
{\rm J. Math. Phys.} {\bf 10}, 1388 (1969).

\bibitem{Affleck}
I. Affleck, T. Kennedy, E. H. Lieb and H. Tasaki,
{\rm Phys. Rev. Lett.} {\bf 59}, 799 (1987),
{\rm Commun. Math. Phys.} {\bf 115}, 477 (1988).

\bibitem{Loa}
I. Loa, S. Gronemeyer, C. Thomsen, R. K. Kremer,
{\rm Solid State Commun.} {\bf 99}, 231 (1996).

\bibitem{correlationlength}
H. L. Swinney and H. Z. Cummins,
{\rm Phys. Rev.} {\bf 171}, 152 (1968).

\bibitem{Halperin2}
L. Van Hove,
{\rm Phys. Rev.} {\bf 95}, 1374 (1954);
B. I. Halperin and P. C. Hohenberg,
{\rm Phys. Rev.} {\bf 177}, 952 (1969).

\bibitem{nishi}
M. Nishi, O. Fujita, J. Akimitsu, K. Kakurai, and Y. Fujii,
{\rm Physica B}{\bf 213}\&{\bf 214}, 275 (1995).

\bibitem{hirota}
K. Hirota, G. Shirane, Q. J. Harris, Q. Feng, R. J. Birgeneau,
M. Hase, and K. Uchinokura,
{\rm Phys. Rev. B}{\bf 52}, 15412 (1995).

\bibitem{Riera2}
J. Riera and S. Koval,
{\rm Phys. Rev. B}{\bf 53}, 770 (1996).

\bibitem{latticeconstant}
J. E. Lorenzo, K. Hirota, G. Shirane, J. M. Tranquada, M. Hase, K. Uchinokura,
H. Kojima, I. Tanaka, and Y. Shibuya,
{\rm Phys. Rev. B}{\bf 50}, 1278 (1994).

\bibitem{latticeconstant2}
Q. J. Harris, Q. Feng, R. J. Birgeneau, K. Hirota, K. Kakurai, J. E. Lorenzo,
G. Shirane, M. Hase, K. Uchinokura, H. Kojima, I. Tanaka, and Y. Shibuya,
{\rm Phys. Rev. B}{\bf 50}, 12606 (1994).

\bibitem{Iio}
T. Nakai, M. Kubota, J. Yu, Y. Inaguma, M. Itoh, T. Kato, and K. Iio.
{\rm Abstracts of the Meeting of the Physical Society of Japan}, 1995
Sectional Meeting, Part 3, p. 187 and private communication with T. Nakai.

\end{references}
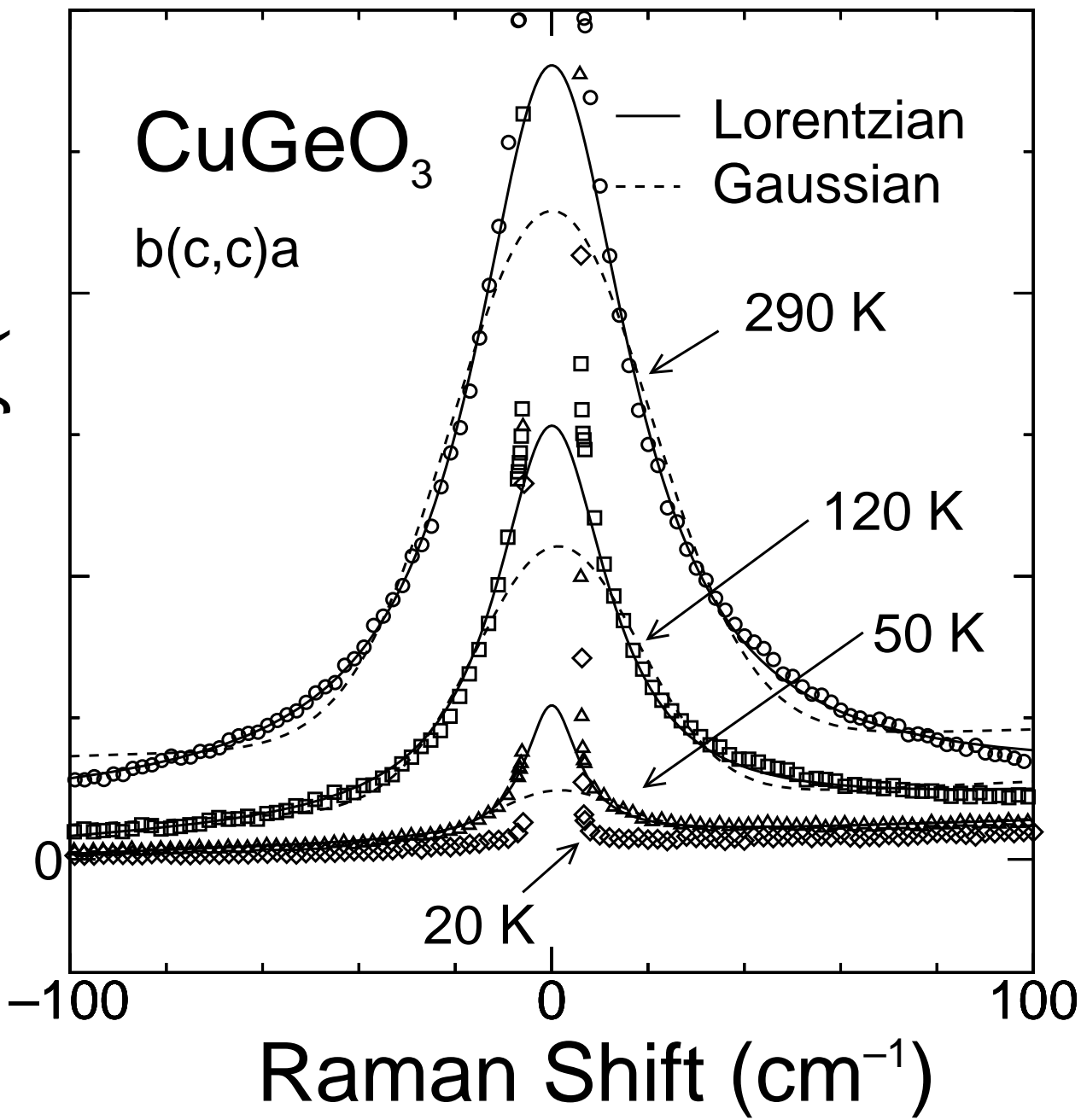
\end{document}